\documentclass[traditabstract,longauth]{aa}  

\usepackage{graphicx}
\usepackage{color}
\usepackage{txfonts}
\usepackage{lscape}

\begin{document}

   \title{The Gaia-ESO Survey. \\ Mg-Al anti-correlation in iDR4 globular
   clusters\thanks{Based on data products from observations made with ESO
   Telescopes at the La Silla Paranal Observatory under programme ID
   188.B-3002.}}

   \titlerunning{Mg-Al anti-correlation in GCs}


   \author{E.~Pancino\inst{\ref{oaa},\ref{asdc}}
        \and
           D.~Romano\inst{\ref{oabo}}
        \and
           B.~Tang\inst{\ref{concepcion}}
        \and
           G.~Tautvai\v sien\. e \inst{\ref{vilnius}}
        \and
           A.~R.~Casey\inst{\ref{cambridge}}
        \and
           P.~Gruyters\inst{\ref{lund}}
        \and
           D.~Geisler\inst{\ref{concepcion}}
        \and
           I.~San Roman\inst{\ref{cefca}}
        \and
           S.~Randich\inst{\ref{oaa}}
        \and
           E.~J. Alfaro\inst{\ref{andalucia}}
        \and
           A.Bragaglia\inst{\ref{oabo}}
        \and
           E.~Flaccomio\inst{\ref{oapa}}
        \and
           A.~J.~Korn\inst{\ref{uppsala}}
        \and
           A.~Recio-Blanco\inst{\ref{nice}}
        \and
           R.~Smiljanic\inst{\ref{poland}}
        \and
           G.~Carraro\inst{\ref{unibo}}
        \and
           A.~Bayo\inst{\ref{valparaiso}}
        \and
           M.~T. Costado\inst{\ref{andalucia}}
        \and
           F.~Damiani\inst{\ref{oapa}}
        \and
           P.~Jofr\'e\inst{\ref{cambridge},\ref{portales}}
        \and
           C.~Lardo\inst{\ref{liverpool}}
        \and
           P.~de~Laverny\inst{\ref{nice}}
        \and
           L.~Monaco\inst{\ref{bello}}
        \and
           L.~Morbidelli\inst{\ref{oaa}}
        \and
           L.~Sbordone\inst{\ref{millennium},\ref{puc}}
        \and
           S.~G.~Sousa\inst{\ref{porto}}
        \and
           S.~Villanova\inst{\ref{concepcion}}
           }

   \institute{INAF-Osservatorio Astrofisico di Arcetri, Largo Enrico Fermi 5,
              50125 Firenze, Italy\\
              \label{oaa}
              \email{pancino@arcetri.inaf.it}
         \and
              INAF-Osservatorio Astronomico di Bologna, Via Ranzani 1, 40127
              Bologna, Italy\label{oabo}
         \and
              ASI Science Data Center, Via del Politecnico snc, 01333, Roma,
              Italy\label{asdc}
         \and
              Departamento de Astronom\'{i}a, Casilla 160-C, Universidad de
              Concepci\'on, Concepci\'on, Chile \label{concepcion}
         \and
              Institute of Theoretical Physics and Astronomy, Vilnius 
              University, Saul\.{e}tekio av. 3, LT-10257 Vilnius, Lithuania
              \label{vilnius}
         \and
              Institute of Astronomy, University of Cambridge, Madingley Road, 
              Cambridge CB3 0HA, United Kingdom\label{cambridge}
         \and
              Lund Observatory, Department of Astronomy and Theoretical 
              Physics, Box 43, SE-221 00 Lund, Sweden\label{lund}
         \and
              Centro de Estudios de F\'\i sica del Cosmos de Arag\'o n (CEFCA),
              Plaza San Juan 1, E-44001 Teruel, Spain\label{cefca}
         \and
              Instituto de Astrof\'{i}sica de Andaluc\'{i}a-CSIC, Apdo. 3004, 
              18080 Granada, Spain\label{andalucia}
         \and
              INAF-Osservatorio Astronomico di Palermo, Piazza del Parlamento
              1, 90134, Palermo, Italy\label{oapa}
         \and
              Department of Physics and Astronomy, Uppsala University, Box 516,
              SE-751 20 Uppsala, Sweden\label{uppsala}
         \and
              Laboratoire Lagrange (UMR7293), Universit\'e de Nice Sophia
              Antipolis, CNRS,Observatoire de la C\^ote d'Azur, CS 34229,F-06304
              Nice cedex 4, France\label{nice}
         \and
              Nicolaus Copernicus Astronomical Center, Polish Academy of
              Sciences,  ul. Bartycka 18, 00-716, Warsaw, Poland\label{poland}
         \and
              Dipartimento di Fisica e Astronomia, Universit\`a di Padova, 
              Vicolo dell'Osservatorio 3, 35122 Padova, Italy\label{unibo}
         \and
              Instituto de F\'isica y Astronomi\'ia, Universidad de 
              Valpara\'iso, Chile\label{valparaiso}
         \and
              N\'ucleo de Astronom\'ia, Facultad de Ingenier\'ia, Universidad 
              Diego Portales,  Av. Ejercito 441, Santiago, Chile\label{portales}            
         \and
              Astrophysics Research Institute, Liverpool John Moores 
              University, 146 Brownlow Hill, Liverpool L3 5RF, United 
              Kingdom\label{liverpool}
         \and
              Departamento de Ciencias Fisicas, Universidad Andres Bello, 
              Fernandez Concha 700, Las Condes, Santiago, Chile\label{bello}
         \and
              Millennium Institute of Astrophysics, Av. Vicu\~{n}a Mackenna 
              4860, 782-0436 Macul, Santiago, Chile\label{millennium}
         \and
              Pontificia Universidad Cat\'{o}lica de Chile, Av. Vicu\~{n}a 
              Mackenna 4860, 782-0436 Macul, Santiago, Chile\label{puc}
         \and
              Instituto de Astrof\'isica e Ci\^encias do Espa\c{c}o, 
              Universidade do Porto, CAUP, Rua das Estrelas, 4150-762 Porto, 
              Portugal\label{porto}
              }

   \date{Received Month DD, YYYY; accepted Month DD, YYYY}

 
  \abstract{We use Gaia-ESO Survey iDR4 data to explore the Mg-Al
  anti-correlation in globular clusters, that were observed as calibrators, as a
  demonstration of the quality of Gaia-ESO Survey data and analysis. The results
  compare well with the available literature, within 0.1~dex or less, after a
  small (compared to the internal spreads) offset between the UVES and the
  GIRAFFE data of 0.10-0.15~dex was taken into account. In particular, we present
  for the first time data for NGC~5927, one of the most metal-rich globular
  clusters studied in the literature so far with [Fe/H]=--0.49~dex, that was
  included to connect with the open cluster regime in the Gaia-ESO Survey
  internal calibration. The extent and shape of the Mg-Al anti-correlation
  provide strong constraints on the multiple population phenomenon in globular
  clusters. In particular, we studied the dependency of the Mg-Al anti-correlation extension
  with metallicity, present-day mass, and age of the clusters, using GES data in
  combination with a large set of homogenized literature measurements. We find a
  dependency with both metallicity and mass, that is evident when fitting for the
  two parameters simultaneously, but no significant dependency with age. We
  confirm that the Mg-Al anti-correlation is not seen in all clusters, but
  disappears for the less massive or most metal-rich ones. We also use our
  dataset to see whether a normal anti-correlation would explain the low
  [Mg/$\alpha$] observed in some extragalactic globular clusters, but find that
  none of the clusters in our sample can reproduce it, and more extreme chemical
  compositions (like the one of NGC~2419) would be required. We conclude that GES
  iDR4 data already meet the requirements set by the main survey goals, and can
  be used to study in detail globular clusters even if the analysis procedures
  were not specifically designed for them.}

   \keywords{Surveys -- Stars: abundances -- globular clusters: general --
   globular clusters: individual : NGC~5927}

   \maketitle
%

\section{Introduction}

The phenomenon of multiple populations in globular clusters (GCs) has been
intensively studied in the last 20-30 years, but we still lack a clear
explanation of its origin \citep{gratton12}. The abundance variations pattern
pinpoints the CNO-cycle burning of hydrogen as the major source of the
phenomenon, because most of the elements that are observed to vary in GCs are
used as catalysts in various CNO sub-cycles, where they are depleted or
accumulated depending on the particular reaction rates. However, a hot debate is
still ongoing on which types of polluters convey the processed material into
the GC insterstellar gas reservoir, and how it is recycled to pollute a fraction
of the GC stars \citep[see][for
references]{dercole08,decressin07,larsen12a,renzini15,bastian15}.

The Mg-Al anti-correlation is of particular importance, because unlike the C-N
and Na-O ones, its extension varies significantly from one GC to the other, to
the point of disappearing completely in some GCs. Mg and Al are involved in the
hot Mg-Al cycle, that requires high temperatures
\citep[$\sim$10$^8$~K,][]{denissenkov15,renzini15} and therefore its study can
place very strong constraints on the type of star that is responsible for the
peculiar chemistry observed in GCs. Another advantage of studying Mg and Al is
that they suffer much less internal mixing compared to C and N, or even Na and
O, therefore the observed abundances do not depend on the evolutionary status of
a star. 

\begin{table*}
\caption{\label{tab:clusters} Basic properties of the GC sample,  listing:
   [Fe/H] and RV from \citet{harris96,harris10}; the present-day mass from
   \citet[][annotated with $^a$]{mclaughlin05} or from \citet[][annotated with
   $^b$]{mandushev91}; the median [Fe/H] and RV from GES data; and the number 
   of GES member stars analyzed. }
\centering          
\begin{tabular}{lcccccc}
\hline\hline       
Cluster & [Fe/H]$_{\rm{H96}}$ & RV$_{\rm{H96}}$ & $\log(M/M_{\odot})$ & 
[Fe/H]$_{\rm{GES}}$ & RV$_{\rm{GES}}$ & N$_{\star}$ \\
        & (dex) & (km~s$^{-1}$) & (dex) & (dex) & (km~s$^{-1}$) & \\
\hline                    
NGC 104 (47 Tuc) & --0.72 &    --18.0 & 6.05$\pm$0.04$^a$ & --0.71$\pm$0.02 &  --17.6$\pm$0.8 & 119 \\
NGC 362          & --1.36 &     223.5 & 5.53$\pm$0.04$^a$ & --1.12$\pm$0.03 &   222.3$\pm$0.6 &  73 \\
NGC 1851         & --1.18 &     320.5 & 5.49$\pm$0.04$^a$ & --1.07$\pm$0.04 &   320.2$\pm$0.5 &  89 \\
NGC 1904 (M 79)  & --1.60 &     205.8 & 5.20$\pm$0.04$^a$ & --1.51$\pm$0.03 &   205.2$\pm$0.5 &  30 \\
NGC 2808         & --1.14 &     101.6 & 5.93$\pm$0.05$^a$ & --1.03$\pm$0.03 &   103.7$\pm$1.4 &  45 \\
NGC 4833         & --1.85 &     200.2 & 5.20$\pm$0.21$^b$ & --1.92$\pm$0.03 &   200.6$\pm$1.0 &  28 \\
NGC 5927         & --0.49 &   --107.5 & 5.32$\pm$0.21$^b$ & --0.39$\pm$0.04 & --102.5$\pm$0.7 &  85 \\
NGC 6752         & --1.54 &    --26.7 & 5.16$\pm$0.21$^b$ & --1.48$\pm$0.04 &  --26.3$\pm$0.7 &  57 \\
NGC 7089 (M 2)   & --1.65 &     --5.3 & 5.84$\pm$0.05$^a$ & --1.47$\pm$0.03 &   --1.8$\pm$1.3 &  46 \\
\hline
\hline
\end{tabular}
\end{table*}

The Gaia-ESO survey \citep[GES,][]{ges1,ges2}, that is being carried out at the
ESO VLT with FLAMES \citep{flames}, observed GCs as calibrators for the
astrophysical parameters (AP) and abundance ratios \citep[][hereafter
P16]{pancino16}. Part of the observed GCs were included in the fourth internal
data release (iDR4), that is based on data gathered from December 2011 to July
2014 and from which the next GES public release will be published through the ESO
archive system\footnote{http://archive.eso.org/cms.html}. The iDR4 data also
include relevant archival data obtained with FLAMES in the GES setups. A
particular advantage of the adopted observing setups is that they allow for an
accurate measurement of the Mg and Al abundance ratios with both the UVES and
GIRAFFE spectrographs, thus providing statistical samples comparable to those
recently obtained by APOGEE \citep{meszaros15} and the FLAMES GC survey
\citep{carretta09a,carretta09b,carretta11,carretta13,carretta14}. 

The paper is organized as follows: in Section~\ref{sec:data} we describe the
data treatment and sample selection; in Section~\ref{sec:results} we present the
results and explore their robustness; in Section~\ref{sec:discussion} we
describe and discuss the behaviour of the Mg-Al abundance variations; in
Section~\ref{sec:conclusions} we summarize our findings and conclusions.


\section{Data sample and treatment}
\label{sec:data}

The GES iDR4 data on GCs are all based on the UVES setup centred around
5800~\AA\ and on the two GIRAFFE setups HR~10 (5339--5619~\AA) and HR~21
(8484--9001~\AA). The selection of calibration targets, which include GCs, was
described in detail by P16. Briefly, 14 GCs were selected to adequately cover
the relevant metallicity range, from [Fe/H]$\simeq$--2.5 to -0.3~dex, 11 of
which were analyzed in iDR4. A few less studied GCs were included at the
beginning of the survey, owing to pointing constraints (see P16 for more
details), and in particular the sample includes NGC~5927, one of the most
metal-rich GCs available. The selection of stars was focussed on red giants,
except in NGC~5927 where mostly red clump stars were selected because of the
high differential reddening and the need to maximize cluster members. Stars
already having GIRAFFE archival observations in the ESO archive were
prioritized, to increase the wavelength coverage by including the GES setups.
Stars already observed with UVES were not repeated. A few fibers were dedicated
to re-observe with UVES some GIRAFFE targets and vice-versa, to allow for
cross-calibration. 

\begin{table*}
\caption{List of the 510 stars that were selected from GES iDR4 as probable 
members and analyzed in this paper. The table is available in its entirety
online and at CDS, here we show a portion to illustrate its contents. The
reported errors are the result of the complex GES homogenization procedure and
thus include random and systematic error sources. 
\label{tab:stars}}
\centering          
\begin{tabular}{lccccccccccc}     
\hline\hline       
CNAME & Cluster & T$_{\rm{eff}}$ & $\delta$T$_{\rm{eff}}$ & log$g$ & $\delta$log$g$ 
& [Fe/H]& $\delta$[Fe/H] & $\log \epsilon_{\rm{Al}}$ & 
$\delta\log \epsilon_{\rm{Al}}$ & $\log \epsilon_{\rm{Mg}}$ & 
$\delta\log \epsilon_{\rm{Mg}}$ \\
 &  & (K) & (K) & (dex) & (dex) & (dex) & (dex) & (dex) & (dex) & (dex) & (dex) \\
\hline                    
12593863-7051321 & NGC4833 & 4673 & 124 & 1.308 & 0.246 & --1.844 & 0.103 & 5.61 & 0.07 & 5.94 & 0.13 \\
13000316-7053486 & NGC4833 & 4675 & 132 & 1.207 & 0.239 & --1.920 & 0.106 & 5.60 & 0.07 & 5.59 & 0.13 \\
12585746-7053278 & NGC4833 & 4678 & 127 & 1.316 & 0.261 & --2.024 & 0.119 & 5.29 & 0.07 & 5.73 & 0.14 \\
12592040-7051156 & NGC4833 & 4623 & 123 & 1.130 & 0.252 & --1.922 & 0.101 & 5.55 & 0.07 & 5.50 & 0.13 \\
12593089-7050304 & NGC4833 & 4613 & 123 & 1.112 & 0.254 & --1.920 & 0.108 & 5.55 & 0.07 & 5.66 & 0.14 \\
12594306-7053528 & NGC4833 & 4635 & 117 & 1.070 & 0.235 & --1.890 & 0.111 & 5.61 & 0.07 & 5.69 & 0.13 \\
\hline                  
\end{tabular}
\end{table*}

All iDR4 data were reduced as described in detail by \citet{sacco14} for spectra
taken with UVES \citep{uves} at high resolution
(R$=\lambda/\delta\lambda\simeq$47\,000) and by \citet{jeffries14} for spectra
taken with GIRAFFE \citep{flames} at intermediate resolution
(R$\simeq$16\,000--20\,000). Briefly, the UVES pipeline \citep{uvespipe} was used
to process UVES spectra, performing the basic reduction steps. Additional data
analysis was performed for UVES with specific software developed at the Arcetri
Astrophysical Observatory. GIRAFFE spectra were processed with a dedicated
software developed at CASU\footnote{http://www.ast.cam.ac.uk/$\sim$ mike/casu/}
(Cambridge Astronomy Survey Unit). 

   \begin{figure}
   \centering
   \includegraphics[width=\columnwidth]{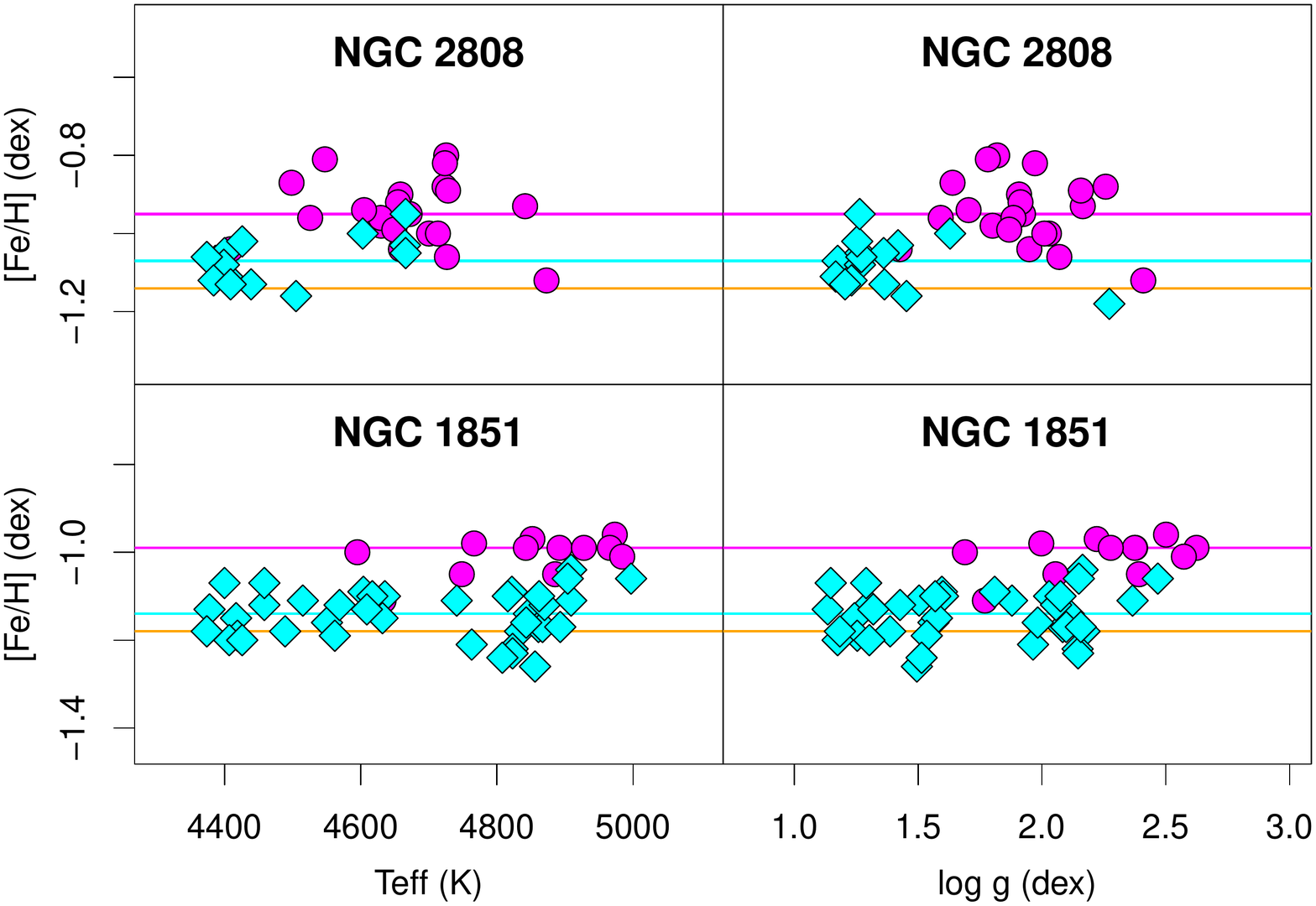}
      \caption{Example of the small (compared to the errors and internal spreads)
      residual offsets in [Fe/H] between UVES and GIRAFFE in GES iDR4 data, in
      two of the sample GCs: NGC~2808 (top panels) and NGC~1851 (bottom panels).
      The left panels show [Fe/H] as a function of T$_{\rm{eff}}$ and the right
      ones of log$g$. UVES stars are plotted as cyan symbols, with their median
      [Fe/H] as a cyan line. GIRAFFE stars are plotted as magenta symbols, with
      their median [Fe/H] as a magenta line. The reference [Fe/H] from
      \citet{harris96,harris10} is plotted as an orange line.} \label{fig:trends}
   \end{figure}

\subsection{Abundance analysis}

The GES abundance analysis of UVES spectra was described in detail by
\citet{smiljanic14} and Casey et al. (in preparation), while the analysis of
GIRAFFE spectra by Recio-Blanco et al. (in preparation). Both are carried out by
many research groups, using several state-of-the-art techniques. Because of the
GES complexity, the data analysis is performed iteratively in each internal data
release (iDR), gradually adding not only new data in each cycle, but also new
processing steps to take into account lessons learned in the previous iDRs
(offsets or trends identified through early science projects) or to increase the
number of elements measured (from molecules, or faint features), or finally by
adding detail to the measurements (corrections for non-LTE, rotational
velocities, veiling, and many more).  This methodology allows for a better
quantification of the internal and external systematics, that are evaluated in a
process of homogeneization of all node results, producing the final GES
recommended APs and abundance ratios, as described by P16 and Hourihane et al.
(in preparation). 

To make the GES data analysis as uniform as possible, the analysis of F, G, and K
type stars relies on a common set of atmospheric models \citep[the MARCS
grid,][]{marcs}, a common linelist \citep{heiter15b}, and -- for those methods
that require it -- a common library of synthetic spectra \citep[computed with
MARCS models and based on the grid by][]{laverny12}. The Solar reference
abundances adopted in this paper were those by \citet{grevesse07}. As mentioned,
iDR4 abundances are computed in the LTE regime, and only future releases will
include non-LTE corrections. Moreover, the GES homogenous analysis relies on a
rich set of calibrating objects (including GCs), selected as described by P16. In
particular, the external calibration of FGK stars in iDR4 relies mostly on the
Gaia benchmark stars \citep{jofre14,blanco14,heiter15a,hawkins16}. 

   \begin{figure*}
   \centering
   \includegraphics[width=\textwidth]{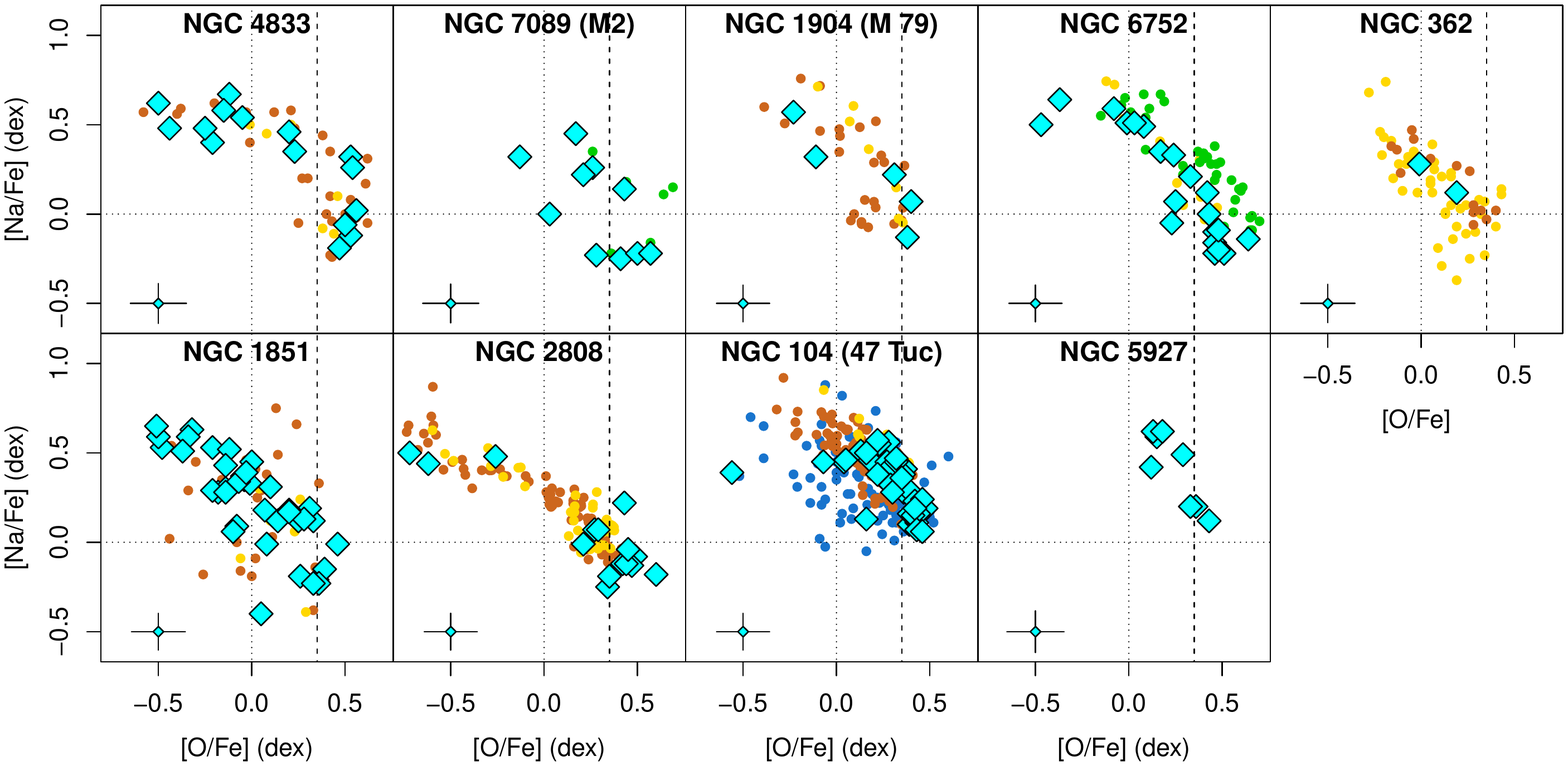}
   \vspace{-4.5cm}
      \caption{Na-O anti-correlation. Panels show different GCs, sorted by
      increasing metallicity from left to right and from top to bottom. Dotted
      lines mark the Solar abundance ratios, dashed lines the typical halo
      $\alpha$-enhancement. GES UVES data are plotted as cyan diamonds;
      literature data from the FLAMES GC survey are plotted in brown for GIRAFFE
      and in gold for UVES; NGC~6752 and M~2 data by \citet{yong05} and
      \citet{yong14} are plotted in green; the 47~Tuc analysis by
      \citet{cordero14} is plotted in blue. Typical (median) errorbars are
      reported on the lower-left corner of each panel.}
       \label{fig:NaO}
   \end{figure*}

When comparing the iDR4 abundances obtained from UVES and GIRAFFE, small (i.e.,
comparable to the internal spreads) offsets in the abundance ratios were found
($\sim$0.10--0.15~dex, depeding on the GC), as shown by P16. For the present
analysis, we reported the [Fe/H] GIRAFFE measurements to the UVES scale using the
difference between the median abundance of the two samples in each GC. We
observed that once the [Fe/H] offsets were corrected this way, there were no
significant residual offsets when comparing the UVES and GIRAFFE measurements of
the other elements considered in this paper. In any case, in the GES cyclic
processing the recommended values of RVs, APs, and chemical abundances generally
improve from one iDR to the next (see Randich et al., in preparation, and P16).
We thus expect the offsets to reduce considerably in future GES releases. Most
importantly, as Figure~\ref{fig:trends} shows, in iDR4 there are no significant
trends of [Fe/H] as a function of T$_{\rm{eff}}$ or log$g$ for either UVES or
GIRAFFE results. 

\subsection{Sample selection}
\label{sec:sele}

We applied the same quality selection criteria of the GES public release (that
will be described in the ESO release documentation) to the iDR4 recommended
results. For the cool giants in GCs, these are:
$\delta$T$_{\rm{eff}}/$T$_{\rm{eff}}<$5\%, $\delta$log$g$<0.3~dex, and
$\delta$[Fe/H]$<$0.2~dex. We also left out all stars that lacked AP or RV
determinations. 

We then selected GC probable members using the median [Fe/H] and RV \citep[as
done by][]{lardo15} as a reference for each GC, and removing all stars that
deviated more than 3\,$\sigma$ from it. As discussed by P16, the GES median
[Fe/H] and RV generally agree with reference literature values
\citep{harris96,harris10}. The members selection was quite straightforward,
because the vast majority of field stars have roughly Solar metallicity and RV
approximately 0$\pm$50~km~s$^{-1}$, thus the GC stars differ significantly from
field stars in at least one of [Fe/H] or RV. 

The above selections lead to highly varying sample sizes for UVES and GIRAFFE
depending on several factors like spectral quality (S/N ratio, spectral defects),
observing conditions (sky, seeing), availability of previous information
(photometry, membership, other archival data), and cluster (crowding, GC
compactness, distance, metallicity). Of the 11 GCs included in iDR4 (see P16, for
the selection criteria of calibrating objects) only 10 contained at least 5 red
giants after the quality and membership selections. Of these, we excluded M~15
because the iDR4 analysis of its very metal-poor spectra did not provide
satisfactory results. The final list of 9 analysed GCs is presented in
Table~\ref{tab:clusters}, along with some relevant properties. The final sample
contained 510 stars (159 with UVES and 351 with GIRAFFE) in 9 GCs, that had Mg or
Al measurements. The stars and their relevant properties are listed in
Table~\ref{tab:stars}. 

   \begin{figure*}
   \centering
   \includegraphics[width=\textwidth]{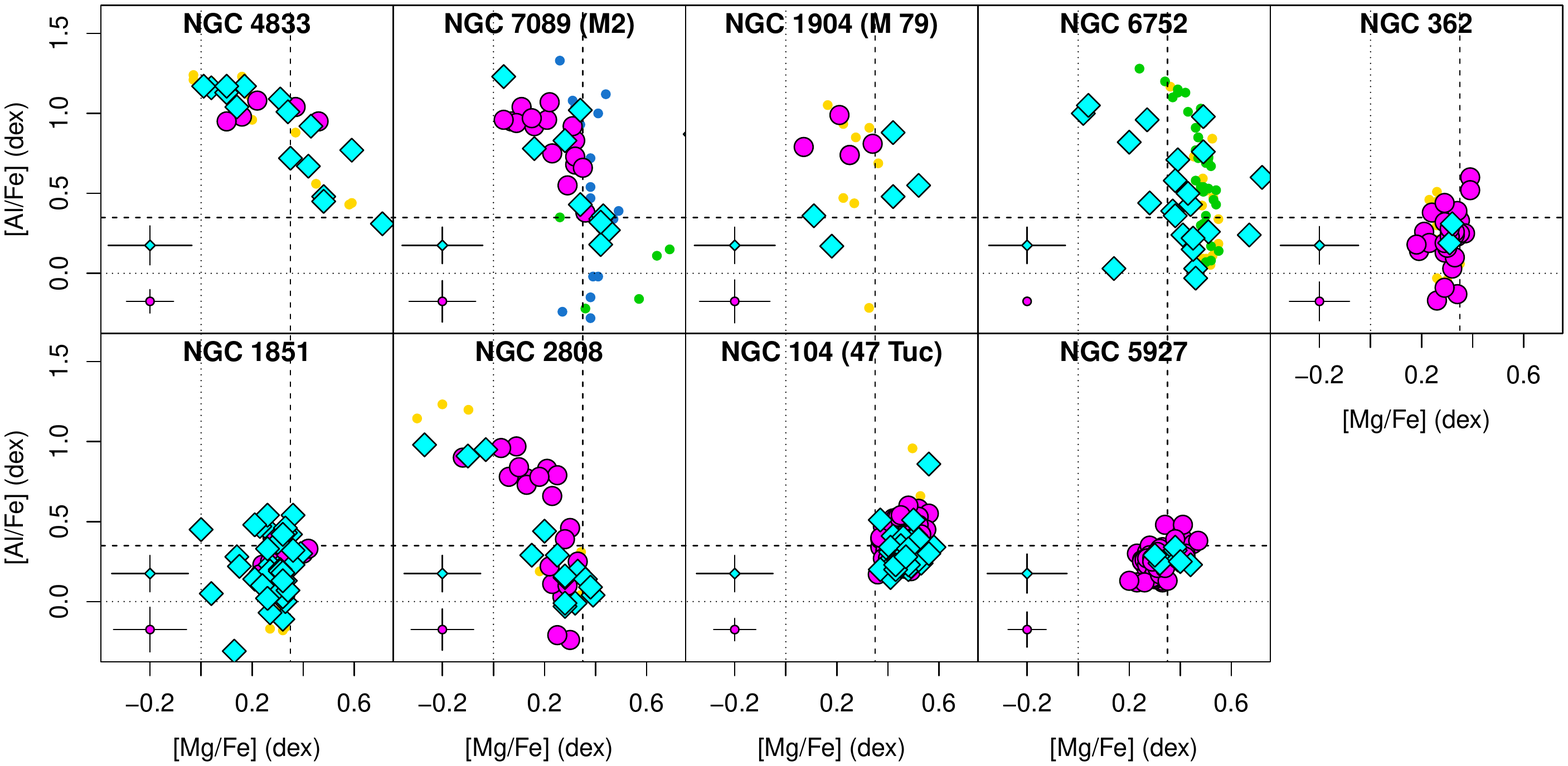}
   \vspace{-4.5cm}
      \caption{Mg-Al anti-correlation. Panels show different GCs, sorted by
      increasing metallicity from left to right and from top to bottom. Dotted
      lines mark the Solar abundance ratios, dashed lines the typical halo
      $\alpha$-enhanced ratios. GES UVES data are plotted as cyan diamonds,
      GIRAFFE data as magenta circles; UVES literature data from the FLAMES GC
      survey are plotted in gold;  NGC~6752 and M~2 data by \citet{yong05} and
      \citet{yong14} are plotted in green; M~2 data by \citet{meszaros15}
      are  plotted in blue. Typical (median) errorbars are reported on the
      lower-left corner of each panel.}
      \label{fig:MgAl}
   \end{figure*}

We stress again that the size and quality of the presented GC sample are
comparable to the two largest GC surveys presented in the literature so far,
i.e., the FLAMES GC survey and the APOGEE sample. 


\section{Results}
\label{sec:results}

\subsection{A quality control test on the Na-O anti-correlation}

We started by comparing our results for the well studied Na-O anti-correlation
with: the FLAMES GC survey by
\citet{carretta09a,carretta09b,carretta11,carretta13,carretta14}; the 47~Tuc
data by \citet{cordero14}; the NGC~6752 study by \citet{yong05}; and the M2
studies by \citet{yong14} and \citet{meszaros15}. We restricted the comparisons
to high-resolution studies (R$>$15\,000) of red giants. The results are plotted
in Figure~\ref{fig:NaO}, where only UVES measurements appear because oxygen is
not included in the GES GIRAFFE setups. 

As can be seen, the GES measurements agree well with the literature ones, in
spite of the different methods, linelist selections, models, and data sets
involved. The median offsets, measured by taking the difference between the
median abundances obtained by GES and in the literature for each GC\footnote{In
many cases, the stars in common between GES and the literature are too few or
missing, therefore we preferred to use the differences between the median of
each sample.}, were in general lower than $\simeq$0.1~dex. We note that for
47~Tuc the GES data are less spread than the literature ones in [O/Fe], but they
do not sample the full extension of [Na/Fe], most probably because of the
quality selection criteria described in Section~\ref{sec:sele}, that penalize
oxygen abundances derived mostly from the weak [O~I] line at 6300~\AA. We
also note that the GES data for NGC~2808 show two well separated clumps of stars
while the literature data apparently display a more continuous distribution. We
ascribe this to our small sample which, being randomly chosen, picked stars near
two most populated peaks of the underlying distribution, which contains five
separate groups \citep{carretta15}. The apparently continuous distribution of
literature data is mostly driven by the GIRAFFE measurements (brown dots), which
are more numerous but less precise than the UVES ones (gold dots).

We present here for the first time [Na/Fe] and [O/Fe] abundance ratios for
NGC~5927, one of the most-metal rich GC studied with high resolution spectroscopy
in the literature so far. NGC~5927 displays the same stubby Na-O anti-correlation
as 47~Tuc, the other metal-rich GC in the sample: while the upper [Na/Fe] limit
is the same as any other GCs, and is governed by the equilibrium abundance of the
NeNa hot cycle, the lowest [Na/Fe] abundances are slightly super-Solar rather
than sub-Solar, as expected for field stars at the same metallicity, as further
discussed in Section~\ref{sec:sample}.

In conclusion, the presented comparison confirms that the atmospheric parameters
resulting from the GES homogenized analysis are well determined (see also P16).

\subsection{Mg-Al anti-correlation}

The Mg-Al anti-correlation for the selected iDR4 stars is plotted in
Figure~\ref{fig:MgAl}, along with the available literature data. In contrast to
the Na-O anti-correlation, we present both UVES and GIRAFFE measurements. Our
measurements compare well with the literature, with small offsets that are
$<$0.1~dex, i.e., within the quoted errors, as in the Na-O case.


For NGC~1904 there are few stars and they appear quite scattered. For the other 8
GCs, however, we clearly see that the Mg-Al anti-correlation has a variable
extension. Four GCs have a well-developed and curved Mg-Al anti-correlation:
NGC~2808, NGC~4833, NGC~6752, and M~2. Two GCs have a stubby Mg-Al distribution:
NGC~362 and NGC~1851 which mostly display an [Al/Fe] spread and no significant
[Mg/Fe] spread. The two most metal-rich GCs in the sample, 47~Tuc and NGC~5927,
show no clear signs of an anti-correlation. This behaviour was already noted by
\citet{carretta09a}, who explicitly mentioned the GC present-day mass and
metallicity as the two main parameters driving the extent of the Mg-Al
anti-correlation (see Section~\ref{sec:discussion} for more discussion on this
point).

We did not detect any significant variation of the combined abundance of Mg
and Al. This is consistent with no net production of these elements, but just
the result of the conversion of Mg into Al during the Mg-Al cycle. Concerning the
Al-Si branch of the Mg-Al cycle \citep[see also][]{yong05,carretta09a}, we looked
for Si variations in our sample, but unfortunately GES iDR4 contains only a few
Si measurements that pass all the criteria employed to select the sample stars.
Inspection of the [Si/Fe] ratio as a function of [Al/Fe] or [Mg/Fe] for the few
stars with reliable Si measurements in iDR4 did not reveal any clear trend.


\section{Discussion}
\label{sec:discussion}

   \begin{figure}
   \centering
   \includegraphics[width=\columnwidth]{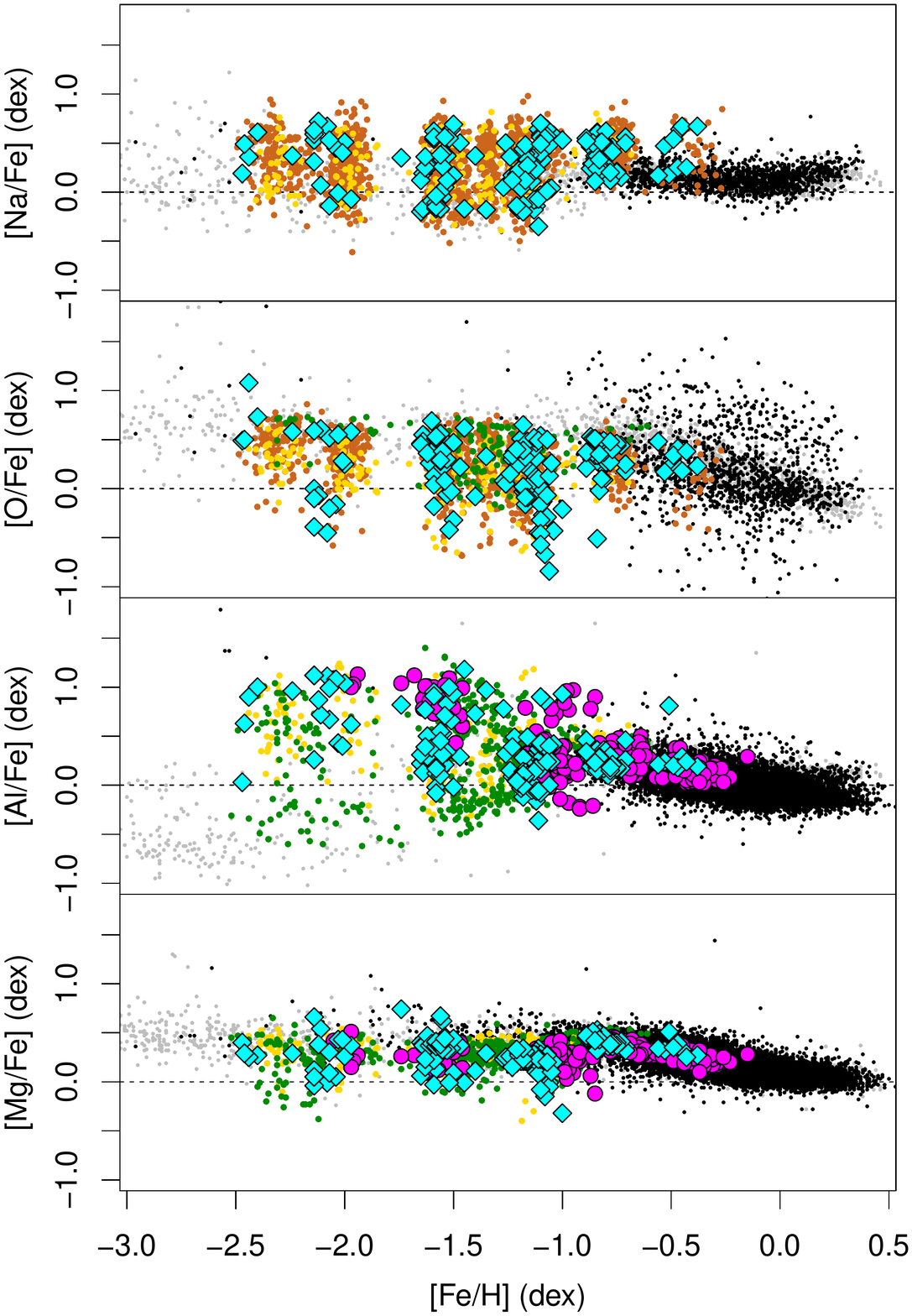}
      \caption{Run of the four main anti-correlated elements as a function of
      [Fe/H]. For MW field stars, GES iDR4 results are plotted as black dots
      and SAGA metal-poor stars as grey dots. Homogenized APOGEE data are
      plotted in green and FLAMES GC surveys data are plotted in yellow for
      UVES and brown for GIRAFFE. GES iDR4 measurements are plotted in cyan for
      UVES and magenta for GIRAFFE.}
      \label{fig:GCfield}
   \end{figure}

To put our results in context, we combined the GES iDR4 data with the FLAMES GC
survey  \citep{carretta09a,carretta09b} and the APOGEE survey \citep{meszaros15}
measurements. Literature data were shifted in both [Fe/H] and the [El/Fe]
abundance ratios by small amounts ($\leq$0.1~dex) to place them on the GES iDR4
scale. The shifts were computed using the median values of key elements for the
GCs in common among studies\footnote{We had M~2 in common with the APOGEE survey
and 6 GCs in common with the FLAMES GC survey (see also Figures~\ref{fig:NaO} and
\ref{fig:MgAl}). The handful of stars in common among the various studies was not
sufficient to compute reliable shifts, and was removed from the sample, retaining
with precedence the GES data, then APOGEE ones, and then the FLAMES GC survey
ones.}. The combined sample contains $\simeq$1\,300 stars in 28 GCs, having both
Mg and Al measurements (or 2\,500 stars if one counts also the stars having Na or
O, but missing one of Mg or Al).

In the next sections, we discuss some of the Mg-Al anti-correlation properties
that were apparent during a preliminary exploration of the combined sample. We
leave the discussion of other elements to the following GES releases, where more
stars, more GCs, and more elements will be available, and the whole GES
intercalibration procedure will be more refined.

\subsection{Comparison with field stars}
\label{sec:sample}

We started by examining the Na-O and Mg-Al anti-correlation as a function of
metallicity, and we compared the available GC measurements with the Milky Way
(MW) field population. Because iDR4 contains mostly MW stars with
[Fe/H]$\geq$--1.0~dex, we added metal-poor stars extracted from the SAGA
database \citep{saga}. Figure~\ref{fig:GCfield} shows the comparisons. Oxygen
measurements in iDR4 are still quite spread out, because they are often based
solely on the weak [O~I] line at 6300~\AA, and they rely on the generally lower
S/N ratio of field star spectra compared to GC stars (see P16 for details), but
the bulk measurements follow the expected trend. In spite of the heterogeneity
of the sample and of our relatively simple homogeneization method, the agreement
among the plotted studies is remarkably good. 

Two important things should be noted at this point. The first is that both GES
and the FLAMES GC survey use similar instrumental setups, wavelength ranges, and
S/N ratios. GES is targeting mostly MW field stars of higher metallicity, while
the FLAMES GC survey was focused on the Na-O anti-correlation. As a result,
neither of these surveys contains many measurements at [Fe/H]$\leq$--1.7~dex, and
in particular, they do not contain many stars with low values of [Al/Fe] or
[Mg/Fe]\footnote{Both GES and the FLAMES GC survey contain several upper limits
in the most metal-poor GCs, that are not plotted in this paper.}, because they
mostly rely on spectral lines that become weak at those metallicities. On the
contrary, APOGEE measurements are obtained with a different wavelength range and
using different features and selection criteria, and therefore that sample
contains many more stars with low Al or Mg, as can be seen in
Figure~\ref{fig:GCfield}. On the other hand, GES data add NGC~5927 to the sample,
extending the [Fe/H] coverage to [Fe/H]=--0.49~dex, while the two previous
systematic studies considered here reached [Fe/H]$\simeq$--0.7~dex with 47~Tuc
and M~71. 

As was noted by others before, the lower boundary of the Na and Al distribution
in GCs is aligned with the typical field star value at any given metallicity.
Similarly, the upper boundary of the O and Mg distribution in GCs is aligned
with the typical field-star $\alpha$-enhancement at any given metallicity. This
supports the idea that the main contributors to the chemistry of {\em normal}
stars in GCs (often called {\em first generation} stars or {\em unenriched}
stars) are mostly SNe II, like for the field stars at the same metallicity,
with SNe~Ia intervening only above [Fe/H]$\simeq$--1.0~dex. 

The abundance of {\em anomalous stars} (often called {\em second generation} or
{\em enriched} stars) is thought to be governed by CNO cycle processing at high
temperatures \citep{kraft94,gratton04}. The extent of Na variations in GC
stars changes slightly with [Fe/H]. This is mostly governed by the lower boundary
variations of [Na/Fe] in GC stars, that follow the field population behaviour as
discussed. The upper boundary -- governed by the equilibrium abundances reached
in the Ne-Na cycle -- shows only moderate variations in our sample, being roughly
at [Na/Fe]$\simeq$+0.6~dex, and contained within $\pm$0.15~dex\footnote{We
remark here that an extremely homogeneous and populous sample would be required
to better quantify this important aspect.}. The extent of [Al/Fe] variations in
GC stars, instead, changes dramatically with [Fe/H] both in the upper and lower
boundaries. While it was suggested that [Fe/H] is not the sole parameter
governing Al variations (see also Section~\ref{sec:ext}), the Al spread clearly
varies with metallicity, from a maximum of  $\Delta$[Al/Fe]$\simeq$1.5~dex and
more below [Fe/H]$\simeq$--1.0~dex, to $\Delta$[Al/Fe]$\leq$0.5~dex above that
metallicity, where the spread become compatible with measurement uncertainties. 

These considerations lead us to believe that the entire sample of 1\,300 stars
should be used when studying the behaviour of the Mg-Al anti-correlation with
GC properties, to increase the parameter coverage and the statistical
significance of the analysis. Figure~\ref{fig:GCfield} is an example of the
striking power of such a sample, and reveals the importance of [Fe/H] as a
driving parameter for the presence and extent of the Mg-Al anti-correlation.

   \begin{figure}
   \centering
   \includegraphics[width=\columnwidth]{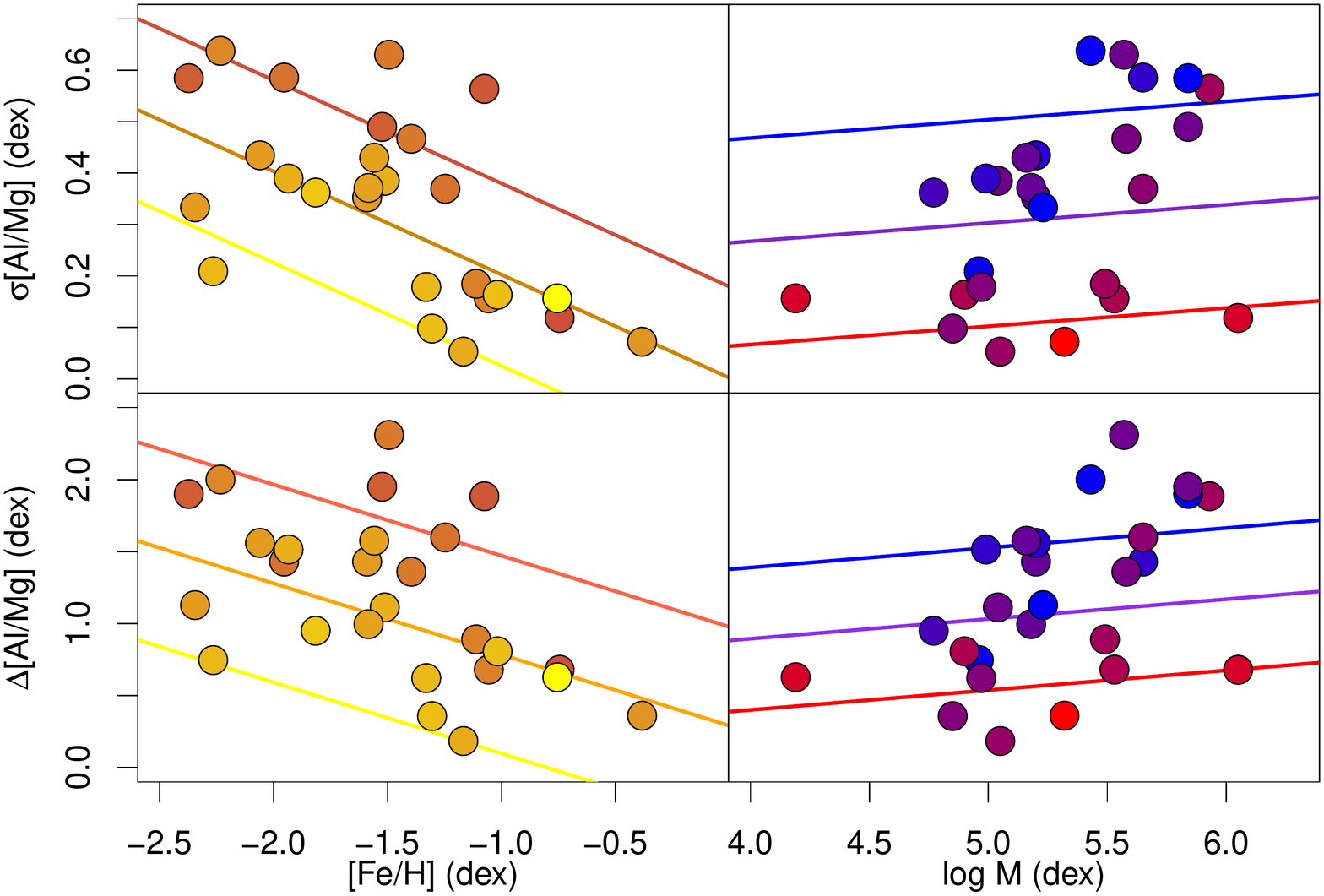}
      \caption{The extent of the Mg-Al anti-correlation, measured as
      $\sigma$[Al/Mg] (upper panels) and $\Delta$[Al/Mg] (lower panels), based on
      the sample described in the text. The behaviour as a function of average
      [Fe/H] (left panels) and total log~M (present-day mass, right panels) of
      each GC is shown. GCs in the left panels are coloured as a function of
      log~M, where yellow corresponds to log~M=4.19~dex (the lowest mass in the
      sample) and dark orange to log~M=6.05~dex (the highest mass). In the right
      panels, points are coloured as a function of their metallicity, with red
      corresponding to [Fe/H]=--0.5~dex (the highest metallicity in the sample)
      and blue to [Fe/H]=--2.5~dex (the lowest metallicity). Our models in the
      form $a$\,[Fe/H]+$b$\,log\,M+$c$ are also plotted as lines coloured based
      on mass or metallicity.} \label{fig:ext}
   \end{figure}

\subsection{Mg-Al anti-correlation extension}
\label{sec:ext}

We have seen that a clear variation of the [Al/Fe] spread with [Fe/H] is apparent
in Figure~\ref{fig:GCfield}, and this is not only caused by the natural [Al/Fe]
variations observed for field stars (the lower [Al/Fe] boundary). The question of
which GC properties govern the extension (or presence) of the Mg-Al
anti-correlation has been explored previously in the literature \citep[see,
e.g.,][for example]{carretta09a,carretta09b,meszaros15,cabrera16}. Both [Fe/H]
and present-day mass were mentioned as the most important parameters in those
works. However, when only [Fe/H] was considered \citep[Figure~4 by][]{cabrera16},
only weak correlations were found, with large spreads and unclear statistical
significance. In that case, 25 GCs were examined with typically 10--20 stars per
GC. Here, we can profit from our combined sample of 28 GCs with $\simeq$50 stars
each on average, as described in Section~\ref{sec:sample}, and re-examine these
parameters as drivers of the Mg-Al anti-correlation. 

We therefore proceeded to fit the data using two different indicators of the
anti-correlation extension, the standard deviation of the [Al/Mg]
distribution and its maximum variation, i.e., the difference between tha maximum
and minimum values of [Al/Mg] for each GC. The two indicators will be expressed
as $\sigma$[Al/Mg] and $\Delta$[Al/Mg] in the following\footnote{We remark
here that both indicators are subject to measurement and statistical biases.
Measurement effects (most notably outliers) tend to produce an overstimate of the
Mg-Al extension, while sampling effects (small sample sizes) tend to produce an
underestimate.}. Figure~\ref{fig:ext} shows the results graphically, where it is
apparent that the most massive GCs tend to have higher values with both
indicators in the plot as a function of [Fe/H], and the most metal-poor ones also
have higher spread in the plot as a function of log~M. If we were to fit the two
parameters separately, we would obtain very high spreads and very weak relations
even with our larger sample.

We therefore employed a linear fit on both parameters simultaneously and we
obtained the following results: 
$$\sigma\rm{[Al/Mg]}=0.19(\pm0.06)~\log M - 0.20(\pm0.05)~\rm{[Fe/H]} -
0.94(\pm0.33)$$
$$\Delta\rm{[Al/Mg]}=0.67(\pm0.21)~\log M - 0.53(\pm0.17)~\rm{[Fe/H]} -
3.16(\pm1.11)$$

The fits are also reported in Figure~\ref{fig:ext}. The p-values of the
$\sigma$[Al/Mg] and $\Delta$[Al/Mg] are 0.0001493 and 0.0005242, respectively,
suggesting that it would be improbable to obtain the observed distribution by
chance (if the chosen model\footnote{Here and in the following, we use the
word {\em model} in the statistical sense, i.e., a way of describing the data
phenomenologically and not a physical model.} was correct). The errors on the
coefficient are also relatively low, suggesting that the two-parameter linear
model is a reasonable description of the data. We thus can conclude that both
parameters\footnote{It is important to stress at this point that no
mass-metallicity relation is apparent in Galactic GCs.} are indeed important in
determining the extension of the Mg-Al anti-correlation, in the sense that we do
find much smaller extensions for GCs that are metal-rich or less massive (or
both). This also supports the results obtained by \citet{carretta10} on the Na-O
data of the FLAMES GC survey, and the photometric analysis carried out by
\citet{milone17}.

This does not mean that the model we adopted is the best one, and it does not
mean that [Fe/H] and logM are the only two parameters at play, especially
considering that the errors on the derived coefficients are of about
$\simeq$30\%, and that the residual distributions, although centered on zero,
have relatively large spreads:
$\rm{med(r.m.s.}_{\Delta\rm{[Al/Mg]}})=-0.005\pm0.768$ and
$\rm{med(r.m.s.}_{\sigma\rm{[Al/Mg]}})=+0.014\pm0.258$. In the present analysis, we have
not used the errors in the fit, because of the heterogenity of the data sources
and therefore of error determinations, but even accounting for that, the
relatively large spreads could point towards some extra parameter. We also tried
a different model, adding a quadratic term in both [Fe/H] and logM but the fit
did not improve significantly. Similarly, when adding the age parameter from
\citet{marin09} or from \citet{van13} as a third linear term, the coefficient was
always low ($<$0.0001), and the quality of the  fit was worse than that of the
two-parameters one. A full statistical analysis of the relation between
anti-correlation parameters and GC properties will be presented in a forthcoming
paper, when the analysis of the whole GES sample of stars in all the observed GCs
will be completed, and we will also have data on the [C/Fe] and [N/Fe] ratios.

The Mg-Al anti-correlation is a problem for the scenarios based on fast rotating
massive stars \citep[FMRS,][]{decressin07} or massive interacting binaries
\citep[MIB,][]{demink09}, which activate CNO burning in their cores but require
very high masses (well above 100~M$_{\odot}$) and some tweaking of the reaction
rates to reproduce the Mg-Al observations. More massive stars would be required
like the super-massive stars \citep[SMS,
$\sim$1000~M$_{\odot}$,][]{denissenkov15}, but these are not observed and
therefore their postulated physics is highly uncertain. We expect a metallicity
dependency for SMS because of the strong wind mass loss \citep{vink11} that would
lead to the formation of smaller SMS at higher metallicity. Asymptotic Giant
Branch polluters (AGB), that activate CNO burning in the shell and also
hot-bottom burning at high masses, can explain naturally the Mg-Al observations,
because both the depletion of Mg and the production of Al are extremely sensitive
to the AGB star metallicity  \citep{ventura16}. However, we remark here that none
of the scenarios presented in the literature so far is entirely free from serious
shortcomings \citep{renzini15}. We also remark that no conclusive answer can be
drawn by considering one anti-correlation only and this, like other works, has to
be considered as a preliminary exploration. 

The correlation of the Mg-Al extent with present-day GC mass has not been
explained in detail in any of the scenarios proposed so far. It would be
necessary to explore whether the observed mass variations among Galactic GCs
(presently in the range 10$^4$--10$^6$~M$_{\odot}$) are sufficient to
significantly change the ability of the forming GCs (with their unknown initial
masses) to retain the polluters ejecta.

\subsection{Low-Mg in extragalactic GCs}
\label{sec:low}

It was reported by various authors \citep{larsen14,colucci14,sakari15} that the
integrated light, high-resolution abundance determinations of extragalactic GCs
tend to have [Mg/Fe] significantly below that of MW GCs, around
[Mg/Fe]$\simeq$0~dex and lower, rather than 0.3--0.4~dex. This observational fact
is difficult to explain with problems in the abundance analysis alone: the
comparison by \citet{colucci16} highlights an underestimate of [Mg/Fe] of
$\simeq$0.2~dex with integrated light spectroscopy for some Galactic GC,
while  Larsen et al. (in preparation) find systematic effects of 0.1~dex at most.
The Mg underabundance is not seen in other $\alpha$-element abundances, that are
consistent with the typical $\alpha$-enhancement expected from metal-poor GCs in
the respective galaxies. In other words, [Mg/$\alpha$] in these metal-poor,
extragalactic GCs is lower than in MW GCs with similar metallicity.

Figures~\ref{fig:MgAl} and \ref{fig:GCfield} show that some Galactic GCs -- not
all -- contain a fraction of stars well below [Mg/Fe]$\simeq$0~dex. The
question then is whether the fraction of low-Mg stars and the Mg spread caused
by a normal Mg-Al anticorrelation would be sufficient to produce an average GC
$<$[Mg/Fe]$>$ close to Solar or even lower, as observed in extragalactic GCs
\citep{larsen16}. While a deeper investigation of this topic is outside the
scope of the present paper, we can use the collected GES and literature samples
to understand if anti-correlations are at least a viable explanation for the
observed low [Mg/$\alpha$] abundances in many extragalactic GCs. In practice, we
averaged the [Mg/$\alpha$] measurements for stars in each GC, which is
appropriate because they are based on relatively weak absorption lines, but can
be an incomplete representation of the abundance in the whole GC and on the
proportions of stars with different Mg content. Integrated light measurements, on
the other hand, represent a complete average -- weighted by star brightness and
cut by limiting magnitude -- of a GC \citep[see][for a comparison between the two
methods]{colucci16}. 

\begin{figure}
   \centering
   \includegraphics[width=\columnwidth]{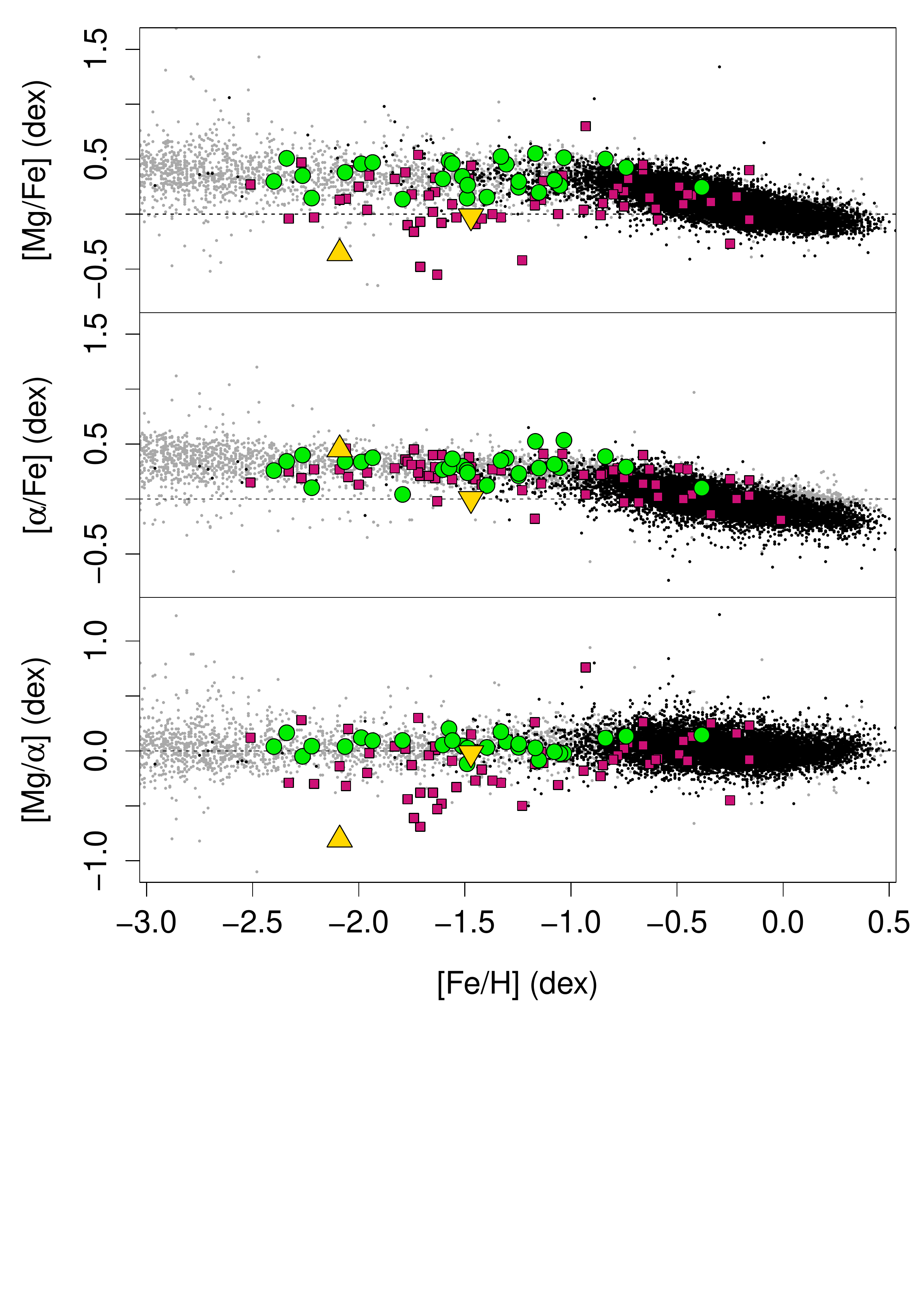}
      \vspace{-3.cm}
      \caption{The average [Mg/Fe], [$\alpha$/Fe], and [Mg/$\alpha$] of our
      collected GES and literature sample of 28 MW GCs (green circles, see
      Section~\ref{sec:discussion}) and of the literature sample of
      extragalactic GCs (purple squares, see Section~\ref{sec:low}). The MW
      reference population is drawn from the GES iDR4 sample (black dots) and
      from the SAGA database of metal-poor stars (grey dots). We also plotted
      NGC~2419 as a yellow upward triangle and Ru~106 as a yellow downward
      triangle.}
      \label{fig:lowMg}
   \end{figure}

We collected literature data on extragalactic GCs in M~31
\citep{colucci09,colucci14,sakari15}, the LMC \citep[Large Magellanic
Cloud,][]{mucciarelli08,
mucciarelli09,mucciarelli10,mucciarelli14,johnson06,mateluna12}, the Fornax
dwarf galaxy \citep{letarte06,larsen12b}, and WLM
\citep[Wolf-Lundmark-Melotte galaxy,][]{larsen14}. To illustrate the effect, we
plotted the data for extragalactic GCs together with the MW field samples and
the Galactic GCs from the collection described in the previous section
(Figure~\ref{fig:lowMg}). The Figure shows the average or integrated abundance
of each GC, where the $\alpha$-elements are represented by Ca and Si, that are
present in all the used studies. As can be noticed, many extragalactic GCs have
normal $\alpha$-enhancement but low [Mg/Fe], and as a result their [Mg/$\alpha$]
ratios are below zero. The MW GCs however, all have [Mg/Fe]$\simeq$0.4~dex --
with very few exceptions -- and have a spread compatible with the errors and the
internal Mg spread of Figure~\ref{fig:GCfield}. 

To our knowledge, the only Galactic GC that contains a sufficient fraction of
stars ($\simeq$50\%) with a sufficiently low [Mg/Fe] is NGC~2419
\citep{mucciarelli12}, reaching as low as [Mg/Fe]$\simeq$--1.0~dex. Based on the
complicated chemistry of NGC~2419, it was suggested that it has extragalactic
origin \citep{mucciarelli12,cohen12,carretta2419,ventura12}, which would fit the
observed data trend. On the other hand Rup~106, which is known to have low
[Mg/Fe] \citep{villanova13}, has a perfectly normal [Mg/$\alpha$], because its
stars are not $\alpha$-enhanced. We conclude that it is difficult to explain the
low integrated [Mg/$\alpha$] values of many extragalactic GCs with the typical
Mg-Al anti-correlation observed in Galactic ones. A more extreme Mg depletion
and a larger fraction of stars with such low Mg would be required, similarly to
what observed in NGC~2419. 

Apart from the extreme morphology of the Mg-Al anti-correlation observed for
example in NGC~2419 (an {\em internal} effect), there is an additional
explanation for the low average [Mg/$\alpha$] of some extragalctic GCs, linked
to the global chemical evolution of their host galaxies (an {\em external}
effect). It has been observed that in dwarf galaxies [Mg/Fe] is lower than the
average $\alpha$-enhancement for stars close to the ``knee" of the [$\alpha$/Fe]
trend. This was explained considering that SNe~Ia produce some amounts of Ca,
Si, and Ti but not Mg, which is produced only by SNe~II \citep{tsujimoto12}. In
that case, we should observe a progressively lower [Mg/$\alpha$] in the field
stars as [Fe/H] increases \citep[as in Figure~10 by][for the
LMC]{mucciarelli12}. The exact distribution would be governed by the global star
formation rate of each galaxy, which governs the metallicity at which the knee
occurs. 

Both the {\em external} and {\em internal} explanations appear viable at the
moment, and they might also operate simultaneously. Further information could be
obtained: (1) by obtaining large and homogeneous samples of field stars with
[Mg/$\alpha$] and [Fe/H] measurements to compare with the available GC
measurements on a galaxy-by-galaxy basis, and (2) by obtaining for the nearest
extragalactic GCs larges sample of individual star abundances.


\section{Summary and conclusions}
\label{sec:conclusions}

We used GES iDR4 data on calibrating globular clusters to explore the Mg-Al
anti-correlation, which is well measured in the GES observing setups and varies
significantly from one GC to the other, and therefore can provide strong
constraints on the GC properties that control the anti-correlation phenomenon. 

Even if iDR4 is a preliminary and intermediate data release, it was the first one
in which many different loops of the internal and external calibration were
closed in the complex GES {\em homogenization} workflow (see P16, Hourihane et
al., in preparation, and Randich et al., in preparation). As result, the
agreement between UVES and GIRAFFE is within the quoted uncertainties, with
0.10--0.15~dex median differences; there are no significant trends of abundance
ratios with the APs, in particular with T$_{\rm{eff}}$ or log$g$; and there are
small offsets with the high-resolution literature data of no more than 0.1~dex.
We also add a new GC, NGC~5927, one of the most metal-rich GCs, that was included
in GES to facilitate the internal calibration in conjunction with open clusters. 

Given the excellent agreement with the literature, we assembled a homogenized
database of $\simeq$1\,300 stars in 28 GCs with [Fe/H], [Mg/Fe], and [Al/Fe]
measurements from GES iDR4, the FLAMES GC survey \citep[][and other papers cited
above]{carretta09a,carretta09b}, and the APOGEE survey \citep{meszaros15}. We
explored two different open topics as a demonstration of the presented data
quality. The first topic concerns the dependency of the Mg-Al anti-correlation
extension with GC global parameters. In particular, it was suggested by
\citet{carretta09a} that the extension depends on both mass and metallicity, but
no formal analysis was performed in that paper owing to the limited sample. The
suspicion was supported by the \citet{meszaros15} data. However a different
analysis by \citet{cabrera16} found a very weak relation between the Mg-Al
extension and [Fe/H] with a large spread and low statistical significance from a
literature database of 20 GC measurements. We profited from our large homogenized
sample, that includes NGC~5927, and we employed a linear fit on cluster mass and
metallicity simultaneously. Our analysis removes any remaining doubt about the
fact the the Mg-Al anti-correlation extension depends on {\em both} mass and
metallicity. Adding age as a third parameter worsened the fit and we concluded
that the Mg-Al anti-correlation does not change significantly with age. 

We also explored another open topic related to the low [Mg/$\alpha$] measured in
some extragalactic GCs \citep{larsen14,colucci14,sakari15}, to see whether a
highly extended Mg-Al anti-correlation could explain the observed trends. We
made the reasonable hypothesis that an average of the available individual star
abundances is comparable with the abundances obtained by integrated light
spectroscopy \citep[see][and references therein]{colucci16}. We concluded that a
normal anti-correlation, no matter how extended, would not reproduce those low
[Mg/$\alpha$] values. A more extreme chemical composition, like that of NGC~2419
\citep{mucciarelli12,cohen12,carretta2419,ventura12} would be required. Besides
this explanation, related to the {\em internal} GC chemical properties, there is
another {\em external} explanation related to the global chemical evolution
properties of the host galaxy and the yields of SNe type Ia and II, but the data
available so far do not allow to discriminate between the two, that could be
either mutually exclusive or cohexist in different GC populations. 

We conclude that the GES data have a quality sufficient to explore the presented
and many other topics related to the chemistry of GCs, providing clear results.
When the whole sample of GCs and of the observed stars will be analyzed,
including also elements that are not completely determined in iDR4, it will be
possible to statistically analyze the entire set of elements that vary in GCs.


\begin{acknowledgements}

   We warmly thank: I.~Cabrera-Ziri for a discussion on the biases in
   determining the extent of the Mg-Al anti-correlation; A.~Mucciarelli for a
   discussion on anomalous GCs like NGC~2419; S.~Larsen for a discussion on the
   phenomenon of low [Mg/$\alpha$] in extragalactic GCs; M.~Gieles for a
   discussion on the possible polluters and their impact on the Mg-Al
   anti-correlation extension; and the referee of this paper, P.~Ventura,
   who offered his insight to improve the manuscript both in its substance and
   form. 

   This research has made use of the following softwares, databases, and online
   resources: topcat \citep{topcat}; the CDS and Vizier databases
   (http://cdsportal.u-strasbg.fr/); the R project (https://www.r-project.org),
   and Rstudio (https://www.rstudio.com/). 
   
   Based on data products from observations made with ESO Telescopes at the La
   Silla Paranal Observatory under programme ID 188.B-3002. These data products
   have been processed by the Cambridge Astronomy Survey Unit (CASU) at the
   Institute of Astronomy, University of Cambridge, and by the FLAMES/UVES
   reduction team at INAF/Osservatorio Astrofisico di Arcetri. These data have
   been obtained from the Gaia-ESO Survey Data Archive, prepared and hosted by
   the Wide Field Astronomy Unit, Institute for Astronomy, University of
   Edinburgh, which is funded by the UK Science and Technology Facilities
   Council. 
   
   This work was partly supported by the European Union FP7 programme through
   ERC grant number 320360 and by the Leverhulme Trust through grant
   RPG-2012-541. We acknowledge the support from INAF and Ministero dell'
   Istruzione, dell' Universit\`a e della Ricerca (MIUR) in the form of the
   grant "Premiale VLT 2012". The results presented here benefit from
   discussions held during the Gaia-ESO workshops and conferences supported by
   the ESF (European Science Foundation) through the GREAT Research Network
   Programme.
   
   M.T.C. acknowledges the financial support from the Spanish {\em Ministerio de
   Econom\'\i a y Competitividad}, through grant AYA2013-40611-P. E.P. and D.R.
   benefited from the International Space Science Institute (ISSI, Bern, CH),
   through funding of the International Team {\em ``The Formation and Evolution
   of the Galactic Halo"}. S.G.S acknowledges the support by Funda\c c\~ ao para
   a Ci\^encia e Tecnologia (FCT) through national funds and a research grant
   (project ref. UID/FIS/04434/2013, and PTDC/FIS-AST/7073/2014). S.G.S. also
   acknowledge the support from FCT through Investigador FCT contract of
   reference IF/00028/2014 and POPH/FSE (EC) by FEDER funding through the program
   {\em ``Programa Operacional de Factores de Competitividade -- COMPETE"}. D.G.,
   B.T., and S.V. gratefully acknowledge support from the Chilean BASAL Centro de
   Excelencia en Astrof\'\i sica y Tecnolog\'\i as Afines (CATA) grant
   PFB-06/2007. L.M. acknowledges support from {\em Proyecto Interno} of the
   Universidad Andres Bello. E.J.A  was supported by Spanish MINECO under grant
   AYA2016-75931-C2-1-P with FEDER funds. R.S. acknowledges support from the
   Polish Ministry of Science and Higher Education (660/E-60/STYP/10/2015).

\end{acknowledgements}

%
%

\bibliographystyle{aa}
\bibliography{MgAl}






   

\end{document}